\documentclass[aps,prb,twocolumn,superscriptaddress,longbibliography]{revtex4-2}
\usepackage{graphicx}
\usepackage{amssymb,amsmath}
\usepackage{bm}
\usepackage{dcolumn}
\usepackage{float}
\usepackage[OT1]{fontenc} 
\usepackage{url}
\usepackage{mathrsfs}
\usepackage{slashed,comment}
\usepackage{color}
\usepackage{verbatim}
\usepackage{enumitem}
\usepackage{soul,physics}
\usepackage[driverfallback=dvipdfm]{hyperref}
\hypersetup{pdfpagemode=FullScreen,colorlinks=true,breaklinks,urlcolor=blue,linkcolor=blue,citecolor=blue}

\usepackage{subfigure}
\usepackage{amssymb}
\usepackage{bm}
\usepackage{graphicx}
\usepackage{amsmath,amssymb}
\usepackage[normalem]{ulem}

\usepackage{pifont}

\begin{document}
 
  \title{Entanglement Growth from Structured Initial States in Many-Body Localized Systems }

  \author{Chen Xu}
  \affiliation{School of Physics, Renmin University of China, Beijing, 100872, China}

  \author{Pengfei Zhang}
  \thanks{PengfeiZhang.physics@gmail.com}
  \affiliation{State Key Laboratory of Surface Physics \& Department of Physics, Fudan University, Shanghai, 200438, China}
  \affiliation{Hefei National Laboratory, Hefei 230088, China}

  \date{\today}

  \begin{abstract}
  Understanding how complex entanglement structures emerge is a central problem in quantum many-body physics. Recent work by Zhang et al. {[Phys. Rev. Lett. \textbf{137}, 020404 (2026)]} has considered structured initial states prepared by evolving a product state under a chaotic Hamiltonian for a finite time before quenching to the target Hamiltonian. In this setup, total entanglement entropy growth in many-body localized systems exhibits two distinct regimes, first increasing and then decreasing as the initial entanglement is tuned. In this work, we identify the physical origin of this behavior by analyzing the dynamics of both the R\'enyi entanglement entropy and the Wehrl-R\'enyi entropy in the random-field XXZ model, the latter of which characterizes multipartite entanglement. We show that a similar non-monotonic dependence on the initial entanglement also appears in the net growth of the Wehrl-R\'enyi entropy for product states polarized along the $z$-direction. The first regime is governed by a finite {local} magnetization associated with local integrals of motion, while the second reflects inter-site correlations. In contrast, for product states in the $x/y$-direction, the entanglement growth exhibits a monotonic decay. Our results provide a more fine-grained picture of how distinct initial-state properties shape entanglement dynamics in many-body localized systems.  
  \end{abstract}
    
  \maketitle

  \section{Introduction} 
  Recent years have witnessed significant breakthroughs in the study of non-equilibrium dynamics in quantum many-body systems. In particular, considerable attention has been devoted to the dynamics of quantum entanglement \cite{guhne2009entanglement,Quantumentanglement2009,nielsen2010quantum,bardarson2012unbounded,serbyn2013universal,kim2014local,de2006entanglement,vznidarivc2008many,nishioka2018entanglement,oliveira2007generic,kim2013ballistic,mezei2017entanglement,bianchi2018linear,nahum2017quantum,von2018operator,calabrese2005evolution,calabrese2007quantum,kawabata2023entanglement,bertini2019entanglement}, which serves as an intrinsic probe of quantumness in generic systems, independent of microscopic details. The typical setup consists of preparing the system in a product state and then evolving it under the desired quantum dynamics, such as Hamiltonian evolution. For generic chaotic systems, quantum thermalization \cite{deutsch1991quantum,srednicki1996thermal,srednicki1999approach,rigol2008thermalization,garrison2018does} predicts that, in the long-time limit, the entanglement entropy (EE) \cite{bennett1996concentrating,eisert2010colloquium,calabrese2004entanglement,kaufman2016quantum,zhou2017operator,nishioka2018entanglement} of a subregion, which measures the bipartite entanglement, approaches the corresponding thermal entropy, which depends only on the energy of the initial state. This effectively washes out most dependence on the detailed structure of the initial state, thereby leaving little imprint of its microscopic features. As a result, the role of initial-state structure in entanglement dynamics remains less explored. 

  On the other hand, there exist various mechanisms that prevent a system from thermalizing. One celebrated example is many-body localization (MBL) \cite{basko2006metal,vznidarivc2008many,pal2010many,altman2015universal,nandkishore2015many,luitz2016extended,zhang2018universal,abanin2019colloquium,serbyn2013local,huse2014phenomenology,altman2012many,schreiber2015observation,choi2016exploring,lukin2019probing,ponte2015many,alet2018many,Prasad2022Initial,Prasad2021ManyBody,Prasad2024SingleParticle}, which originates from strong disorder. In such systems, local integrals of motion (LIOMs) emerge \cite{serbyn2013local,huse2014phenomenology,altman2012many,ros2015integrals,chandran2015constructing}, which protect the initial information even at long times. MBL systems exhibit a characteristic logarithmically slow growth of entanglement~\cite{bardarson2012unbounded,serbyn2013universal,kim2014local,nanduri2014entanglement,toniolo2025dynamical,toniolo2024dynamics}, understood from both LIOM phenomenology and logarithmic lightcones~\cite{deng2017logarithmic,fan2017out,huang2017out,chen2016universal,chen2017out,swingle2017slow,he2017characterizing}. Other examples include non-interacting Hamiltonians, simple quantum circuits such as SWAP circuits \cite{piroli2020exact,bertini2020operator1,bertini2020operator2,reid2021entanglement}, and dynamics with repeated measurements \cite{li2018quantum,skinner2019measurement,chan2019unitary}. These non-thermal systems provides a natural playground to the study of initial-state dependence of the entanglement dynamics. A pioneering work in this direction is by Zhang et al. \cite{zhang2025entanglement}, which focuses on initial states prepared by evolving a product state for a finite duration. Interestingly, it shows that net increase of von Neumann EE in many-body localized systems exhibits two distinct regimes, first increasing and then decreasing as the initial entanglement is tuned. Nevertheless, how these signatures are related to the emergence of local integrals of motion remains unexplored.
  \begin{figure}[t]
    \centering
    \includegraphics[width=0.95\linewidth]{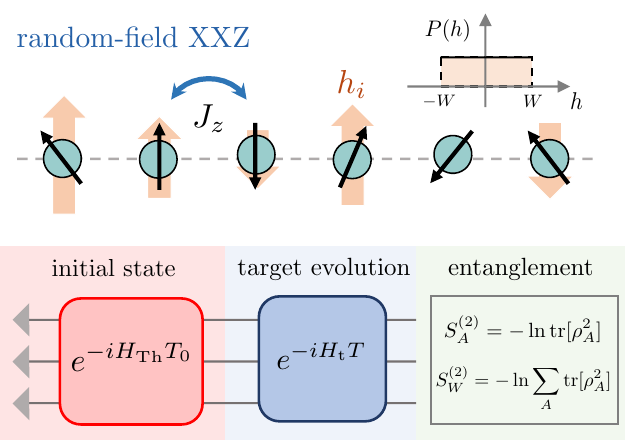}
    \caption{Schematic illustration of our setup. We consider the random-field XXZ model. In particular, we focus on structured initial states prepared by evolving a product state under a chaotic Hamiltonian $H_{\text{Th}}$ for a finite time $T_0$, followed by a quench to the target Hamiltonian $H_t$. We then compute both the entanglement entropy and the Wehrl-R\'enyi entropy. }
    \label{fig:schematics}
  \end{figure}

  In this work, we investigate this problem by analyzing the dynamics of the R\'enyi EE and the Wehrl-R\'enyi entropy (WRE)~\cite{zhang2025exact} in the random-field XXZ model, as illustrated in Fig.~\ref{fig:schematics}. By definition, the WRE is invariant under permutations of spins and thus remains unchanged under SWAP operations. For product states polarized along the $z$-direction, we find that the net growth of the WRE exhibits a similar non-monotonic behavior as the initial-state entanglement is tuned. Using a phenomenological model for MBL systems, we show that this novel behavior can be understood within a generalized Gibbs ensemble description based on LIOMs. In the first regime, the magnetization of LIOMs decreases, effectively enlarging the accessible Hilbert-space dimension and thereby enabling stronger entanglement growth. {As the local magnetization approaches zero}, the system enters a second regime dominated by inter-site correlations. This theoretical picture further suggests the absence of the first regime for product states polarized along the $x$- or $y$-direction, as explicitly confirmed by numerical simulations. These results reveal how local memory of the initial state governs entanglement growth in MBL systems.

  \section{Model \& Setup}
  We focus on quantum spin-$1/2$ systems in one spatial dimension of even length $L$ with open boundary conditions. The Hamiltonian of interest is the random-field XXZ model \cite{abanin2019colloquium,alet2018many}, which reads:
  \begin{equation}
  \hat{H} = \sum_{i=1}^{L-1} \left( \hat{S}_i^x \hat{S}_{i+1}^x + \hat{S}_i^y \hat{S}_{i+1}^y + J_z \hat{S}_i^z \hat{S}_{i+1}^z \right) + \sum_{i=1}^{L} h_i \hat{S}_i^z,
  \end{equation}
  where $\hat{S}_i^\alpha$ with $\alpha \in \{x, y, z\}$ are the spin-$1/2$ operators at site $i$. For simplicity, we take the coupling in the $x$ and $y$ directions as the energy unit and fix $J_z = 0.5$ throughout the manuscript. The on-site magnetic fields $h_i$ are independent random variables drawn uniformly from the interval $[-W, W]$. The system exhibits conservation of the total spin along the $z$-direction, which, under the Jordan–Wigner transformation \cite{fradkin2013field}, is equivalent to particle-number conservation for fermions in a random potential.

  At small $W$, the system is in the thermalized phase, where local information is rapidly scrambled across the entire system and cannot be reconstructed from local measurements \cite{Kitaev2015,maldacena2016remarks,kitaev2018soft,mi2021information,zhang2023information}. As $W$ increases, the disorder becomes stronger, eventually driving the system into a many-body localized phase for sufficiently large $W$, where local conserved quantities $\hat{Q}_i$, known as LIOMs \cite{serbyn2013local,huse2014phenomenology,altman2012many}, emerge. These LIOMs protect local information even at long times, preventing thermalization into conventional thermal ensembles. In the strong-disorder limit $W \to \infty$, these LIOMs coincide with the spin operators $\hat{S}_i^z$, while for large but finite $W$, they acquire a local dressing over a correlation length $\xi$, as indicated by perturbation theory in $1/W$. The conservation of LIOMs implies that the effective dynamics can be captured by a phenomenological model \cite{serbyn2013local,huse2014phenomenology,altman2012many}:
  \begin{eqnarray}\label{Heff}
  \hat{H}_\text{eff} = \sum_{i=1}^L g_i \hat{Q}_i+\sum_{i<j}  J_{ij} \hat{Q}_i \hat{Q}_{j}+\cdots
  \end{eqnarray}
  where $J_{ij} = \tilde{J}_{ij} e^{-|i-j|/\xi}$ is the strength of the two-body interactions between LIOMs, which typically decays exponentially with their separation. The term $\cdots$ denotes possible higher-body contributions that are kept implicit. $\tilde{J}_{ij}$ and $g_i$ are typically modeled as random variables uniformly distributed in the intervals $[-J_0, J_0]$ and $[-g_0, g_0]$, respectively. In this work, when we employ Hamiltonian evolution by the phenomenological model, we (1) replace $\hat{Q}_i = \hat{S}_i^z$, and (2) neglect all higher-body terms. {We also fix $J_0=\xi=1$ and set $g_0=0$, which does not alter our main conclusions within the parameter range tested, as demonstrated in the Appendix~\ref{parameter}.}
  
  We are interested in the entanglement dynamics of the system with structured initial states~\cite{zhang2025entanglement}. {To prepare these states, we start from a product state $|\psi_0\rangle$, randomly chosen from the $z$-basis product states within the half-filling sector, i.e., satisfying $\sum_i \hat{S}_i^z=0$.} The system is then evolved under a Hamiltonian in the thermal phase, fixed as $\hat{H}_{\text{Th}} = \hat{H}(W = 0.5)$, for a duration $T_0$. The resulting state is given by $|\psi_{\text{ini}}\rangle= e^{-i\hat{H}_{\text{Th}}T_0}|\psi_0\rangle$. {This state exhibits preparation-time-dependent local polarizations, connected spin correlations, as well as bipartite and multipartite entanglement. As a result, we refer to the state $|\psi_{\text{ini}}\rangle$ as ``structured''.} It is taken as the initial state for the evolution governed by a second target Hamiltonian $\hat{H}_t$. We consider three different choices for the target Hamiltonian: (I) $\hat{H}_t = \hat{H}(W = 8)$ in the localized phase, (II) the phenomenological model $\hat{H}_t = \hat{H}_{\text{eff}}$, and (III) $\hat{H}_t = \hat{H}(W=0.5)$ in the thermal phase, with an independent realization of the random fields. The state evolved over an additional time duration $T$ is given by
  \begin{equation}
  |\psi(T)\rangle= e^{-i\hat{H}_{t}T}|\psi_{\text{ini}}\rangle= e^{-i\hat{H}_{t}T}e^{-i\hat{H}_{\text{Th}}T_0}|\psi_0\rangle.
  \end{equation}

  To characterize the entanglement of the state $|\psi(T)\rangle$, we partition the system into two subregions, $A$ and $B$. The reduced density matrix of $A$ is given by tracing out qubits in $B$ as $\rho_A=\text{tr}_B[|\psi(T)\rangle \langle \psi(T)| ]$. The (second) R\'enyi EE for the region $A$ is defined as $S^{(2)}_A=-\ln \text{tr}[\rho_A^2]$. We mainly focus on two classes of subregions:
  \begin{enumerate}
  \item The half-chain R\'enyi EE (HCEE) $S^{(2)}_{\text{HC}}$, defined by choosing $A$ to include qubits $i \in \{1,2,\cdots, L/2\}$.

  \item The $n$-qubit R\'enyi EE ($n$QEE) $S^{(2)}_{n\text{Q}}$ is defined as the average of $S^{(2)}_A$ over all \textit{contiguous} $n$-qubit subsystems.
  \end{enumerate}
 In addition, we introduce the Wehrl-R\'enyi entropy (WRE)~\cite{zhang2025exact}, which is equivalent to concentratable entanglement~\cite{PhysRevLett.127.140501} for spin-$1/2$ systems. It quantifies the complexity~\cite{zhang2025exact,xu2024wehrl} and multipartite entanglement~\cite{PhysRevLett.127.140501} of quantum many-body states. The original definition of the WRE $S^{(2)}_W$ is given in terms of the second R\'enyi entropy of the Husimi function \cite{husimi1940some,xu2025bayesian} for quantum spin systems. For simplicity, we adopt an alternative definition directly in terms of $S_A^{(2)}$, as established in~\cite{schenk2007relation} for generic spin-$1/2$ quantum states:
  \begin{equation}\label{eq:defSW}
  S_W^{(2)}= -\ln \bigg(\sum_A e^{-S^{(2)}_A}/C_0\bigg)=-\ln \bigg(\sum_A \text{tr}[\rho_A^2]/C_0\bigg).
  \end{equation}
 Here, $C_0=(6\pi)^L$ is the normalization factor for the equal-weight average; meanwhile the summation runs over all $2^L$ possible choices of the subsystem $A$, including the empty set. 

  In Ref.~\cite{zhang2025entanglement}, two elementary processes of entanglement dynamics were identified: ``build'', which corresponds to the creation of new entanglement, and ``move'', which corresponds to the redistribution of entanglement. For example, a random two-qubit gate can generate new entanglement between qubits, whereas SWAP operations only exchange qubit labels and therefore merely redistribute existing entanglement. From the definition in Eq.~\eqref{eq:defSW}, the WRE remains invariant under SWAP operations. Therefore, it provides a natural way to distinguish between the ``build'' and ``move'' processes . In contrast, both choices of the R\'enyi EE are sensitive to local ``move'' operations and therefore include both contributions. 
     
  \begin{figure}[t]
    \centering
    \includegraphics[width=1\linewidth]{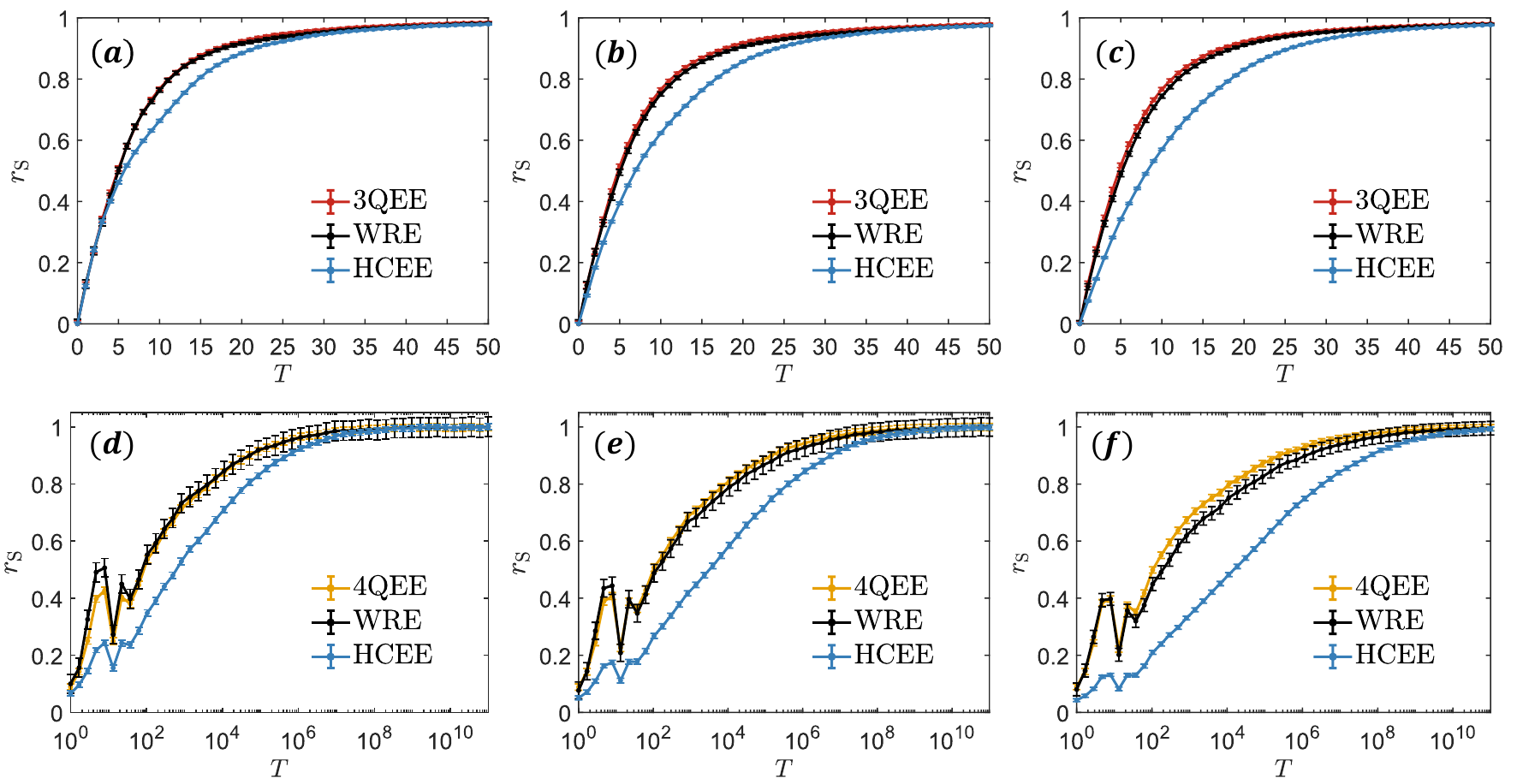}
    \caption{The evolution of the HCEE, $n$QEE, and WRE starting from the initial state prepared with $T_0 = 4.5$. (a-c) Dynamics in the thermalized phase ($W = 0.5$) for system sizes {$L=12$, $14$, and $16$}, respectively. (d-f) Dynamics under the MBL Hamiltonian ($W = 8$) for system sizes {$L=12$, $14$, and $16$}, respectively. { All results are obtained by averaging over 500 independent realizations of the random fields.} }
    \label{fig:num1}
  \end{figure}

  \section{Evolution of Entanglement}
  As time $T$ increases, entanglement in the system builds up rapidly. For systems in the thermalized phase (case III), it has been established that energy conservation leads to diffusive spreading of entanglement, resulting in a sub-ballistic growth of the Rényi EE, $S_{\text{HC}}^{(2)} \propto \sqrt{t}$~\cite{huang2020dynamics,PhysRevLett.122.250602}. {The saturation of the HCEE requires the buildup of correlations between qubits separated by a distance of order $O(L)$, which provides an estimate of the saturation time $\sim L^2$.} In contrast, the $n$QEE (for finite $n$) typically saturates at a finite time that is nearly independent of the system size $L$. For systems in the localized phase (cases I and II), entanglement grows only logarithmically as $S_{\text{HC}}^{(2)} \propto \ln t$, and the saturation of the HCEE requires an exponentially long timescale of order $O(e^L)$, consistent with the absence of efficient information transport. All expectations are consistent with the numerical results presented in Fig.~2, which are obtained by fixing $T_0 = 4.5$ for system sizes {$L=12$, $14$, and $16$}. Here, we normalize the results by introducing $r_{\text{HC}} = \frac{S^{(2)}_{\text{HC}}(T) - S^{(2)}_{\text{HC}}(0)}{S^{(2)}_{\text{HC}}(\infty) - S^{(2)}_{\text{HC}}(0)}$, and analogously for other entanglement measures to facilitate a direct comparison of their dynamical behaviors.

  The evolution of the WRE remains less explored in both thermalized and localized phases. Interestingly, the numerical results in Fig.~\ref{fig:num1} show that the evolution of the WRE closely follows that of the $n$QEE, with $n=3$ in the thermalized case and $n=4$ in the localized case, indicating that the WRE effectively saturates on an $O(1)$ timescale. This can be understood as follows: The WRE involves a summation over all possible choices of subregions. Without the constraint of locality, a typical subregion consists of disconnected domains of subsystem $A$ embedded within the environment $B$. In this picture, the saturation time of the WRE is controlled by the typical size of these domains, which sets the relevant entanglement build-up scale. Assuming that a given site $i$ is the first site in a domain of subsystem $A$, the probability that its nearest-neighbor site $i+1$ also belongs to $A$ is $1/2$. This simple estimate yields a typical domain-wall width of three sites, which is already consistent with the numerical observations. In addition, we highlight that the observed correspondence between the $n$QEE and the WRE further implies that, at least within this time regime, entanglement growth is always governed by the “build” process, independent of whether the system is in the thermalized or localized phase.

  \begin{figure}[t]
    \centering
    \includegraphics[width=1\linewidth]{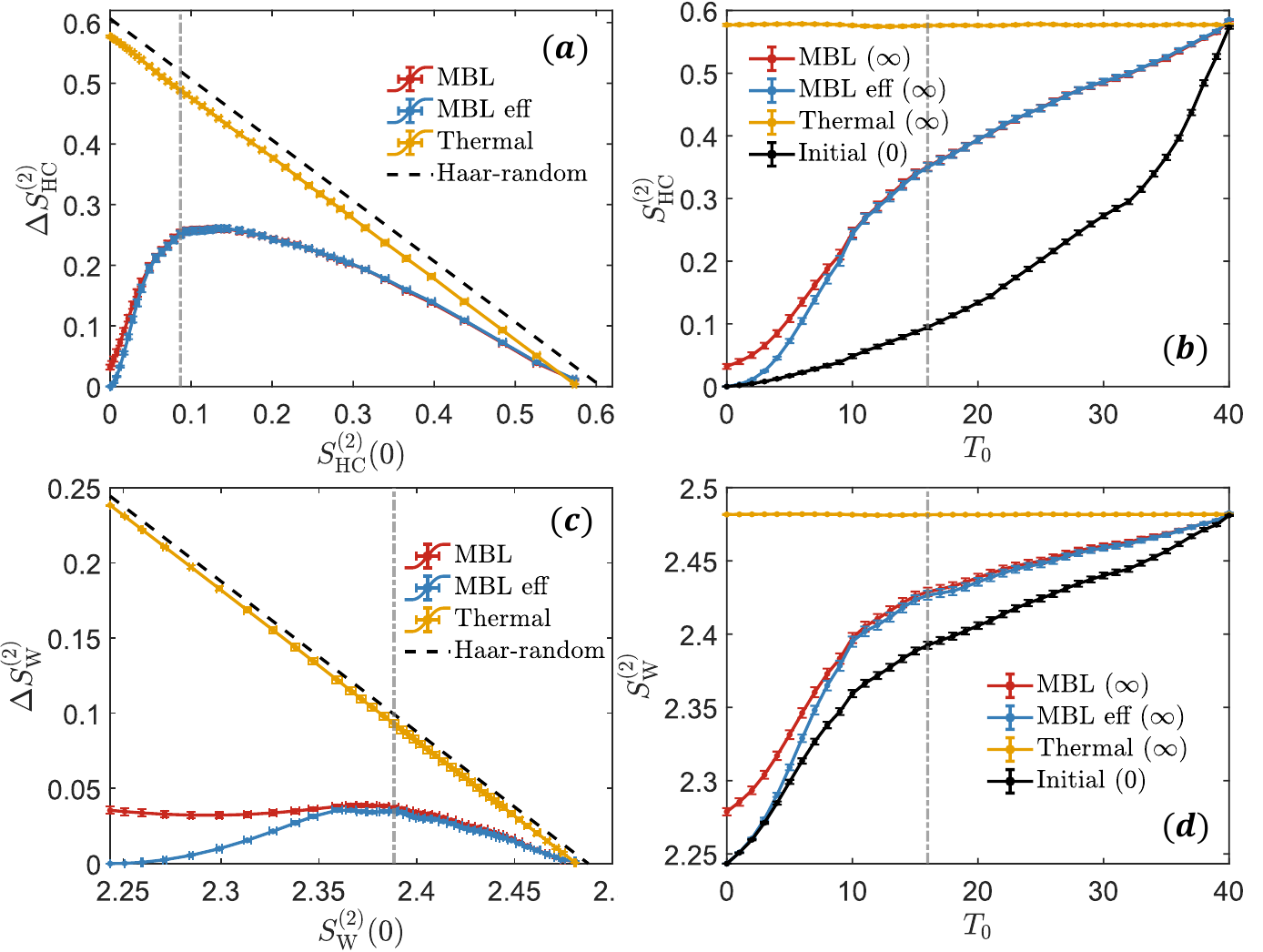}
    \caption{The total entanglement growth of both the HCEE and the WRE is shown for different Hamiltonians and initial states at system size {$L = 16$}. The initial-state entanglement is tuned by varying $T_0$. In the numerics, the saturation value is estimated by evolving to a long time $T = 10^{12}$. The black dashed line indicates the result for Haar-random states, while the gray dot-dashed lines serve as guides to the eye, separating two distinct regimes. {All results are obtained by averaging over 100 independent realizations of randomly sampled product states $|\psi_0\rangle$ and random fields.} }
    \label{fig:num2}
  \end{figure}

  \section{Total Entanglement Growth}
  Next, we focus on the amount of the total entanglement growth for different initial states parametrized by $T_0$ in the limit $T \rightarrow \infty$, defined as $\Delta S^{(2)}_{\text{HC}}= S^{(2)}_{\text{HC}}(\infty) - S^{(2)}_{\text{HC}}(0)$, and analogously for the WRE, $\Delta S^{(2)}_{W}$ \footnote{Following the Ref.~\cite{zhang2025entanglement}, we consider $T_0\in\{0.0,0.25,0.375,0.5,0.625,0.75,0.875,1.0,1.125,1.25,1.5,\\1.75,2.0,2.25,2.5,2.75,3.0,3.3,3.6,3.9,4.2,4.5,5.0,5.5,6.0,\\6.5,7.0,7.5,8.0,8.5,9.0,9.5,10.0,11.0,12.2,13.7,15.7,19.0,\\24.0,32.0,500.0\}$, which ensures a broad and diverse range of initial-state entanglement.}. {
  	For convenience, in this section all entropies are normalized by
  	the corresponding number of sites, namely
  	$S^{(2)}_{W}/L$ and $S^{(2)}_{\mathrm{HC}}/(L/2)$, respectively.
  	Unless otherwise stated, we use the same symbols
  	$S^{(2)}_{W}$ and $S^{(2)}_{\mathrm{HC}}$ below to denote these normalized quantities.
  } The numerical results are presented in the Fig.~\ref{fig:num2}. For Hamiltonian in the thermalized phase, the long-time evolved state exhibits quantum thermalization that only depends on the energy density of the initial state. For $T_0 = 0$, $|\psi_\text{ini}\rangle = |\psi_0\rangle$ is a random product state at half-filling, with energy expectation value $\overline{\langle \psi_0 | \hat{H}_t | \psi_0 \rangle} = 0$, corresponding to an infinite-temperature state. Here, the overline denotes the disorder average over random fields. Furthermore, we expect that evolution under $\hat{H}_{\text{Th}}$ does not lower the effective temperature. Therefore, the system is expected to thermalize to infinite temperature for arbitrary $T_0$, and the entanglement growth for both HCEE and WRE should decrease linearly with the initial entanglement, as demonstrated numerically. The saturation value can be approximated using Haar-random states, as indicated by the black dashed lines, which agree well with the numerical results.

  In contrast, the MBL system does not thermalize to a conventional Gibbs ensemble due to the emergence of LIOMs. The net growth of HCEE in both the microscopic XXZ model and the phenomenological model exhibits non-monotonic behavior, first increasing and then decreasing. This is consistent with the observations in \cite{zhang2025entanglement}, despite the difference between using the von Neumann EE and the R\'enyi EE. The separation between two different regimes are indicated by the gray dot-dashed lines. In addition, the results from the phenomenological model become nearly indistinguishable from those of the microscopic model in the second regime, while small deviations are present in the first regime. We further analyze the growth of WRE, as shown in Fig.~\ref{fig:num2} (c-d). When the total HCEE growth increases, the WRE growth in the phenomenological model also increases, while that in the microscopic model is significantly slower. Near the transition point, the two curves coincide and then begin to decrease as the HCEE growth diminishes. 

  This reveals the origin of the two distinct regimes. The difference between the microscopic model and the phenomenological model arises from the distinction between the LIOMs used in the phenomenological model, $\hat{S}^z_i$, and the microscopic LIOMs, which include additional renormalization from perturbation theory. Let us first focus on $T_0=0$. The initial state is an eigenstate of the phenomenological model, but not for the XXZ model. We can mimic the XXZ evolution using the phenomenological model by applying a finite-depth quantum circuit $\hat{U}_p$ applied to the initial state before evolution under 
  the phenomenological Hamiltonian; this circuit implements the perturbative change of basis. Equivalently, in terms of the microscopic spin operators, this leads to a relaxation of the local magnetization $m_i = \langle \hat{S}^z_i \rangle$. Now we consider finite $T_0$. In the microscopic model, the relaxation of $m_i$ receives contributions from both $\hat{U}_p$ and $e^{-i\hat{H}_t T}$, whereas in the phenomenological model it arises solely from $e^{-i\hat{H}_t T}$. The distinction between the two scenarios becomes negligible when the evolution under the target Hamiltonian alone fully relaxes the local magnetization, identified as the transition point. Therefore, the first regime is governed by the local relaxation of LIOMs. Physically, this yields an increase in the accessible Hilbert-space dimension, which enhances entanglement growth.

  \begin{figure}[t]
    \centering
    \includegraphics[width=1\linewidth]{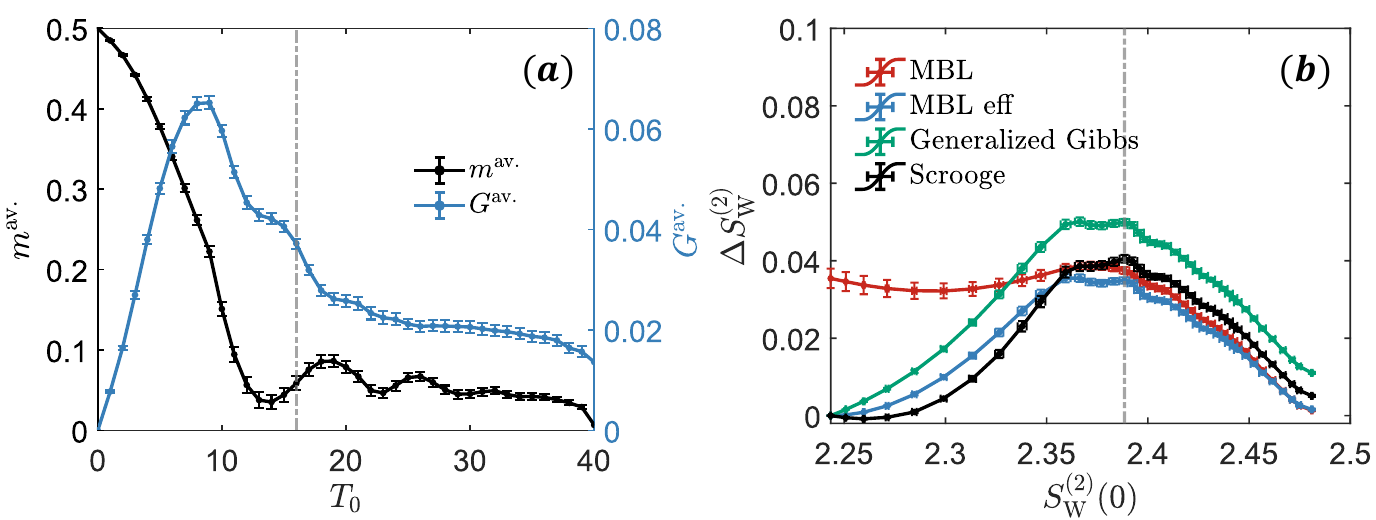}
    \caption{ (a) The averaged local magnetization ${m}^{\text{av}.}$ and nearest-neighbor correlation ${G}^{\text{av}.}$ in the initial state $|\psi_\text{ini}\rangle$ for different preparation times $T_0$ at system size {$L = 16$}. (b) A comparison between numerical results and the predictions of the generalized Gibbs ensemble and the Scrooge random-state ensemble ({ averaged over 100 independent Haar-random states}) for the net growth of the WRE at different $T_0$. {The parameters and regime boundaries are the same as those in Fig.~\ref{fig:num2}.} }
    \label{fig:num3}
  \end{figure}


  	
 	As a verification, we compute the local magnetization of the initial state $|\psi_{\text{ini}}\rangle$ for different values of $T_0$. {We characterize the residual local-spin memory by introducing the averaged local magnetization
 	\begin{equation}
 		m^{\mathrm{av.}}
 		=
 		\frac{1}{L}
 		\sum_j
 		\eta_j
 		\langle\psi_{\mathrm{ini}}|
 		\hat S_j^z
 		|\psi_{\mathrm{ini}}\rangle ,
 	\end{equation}
 	where the factor
 	$
 		\eta_j
 		=
 		2\langle\psi_0|\hat S_j^z|\psi_0\rangle
 		=
 		\pm1
 	$
 	aligns the sign of the local magnetization at each site with that of the
 	initial reference product state $|\psi_0\rangle$.
 	The quantity $m^{\mathrm{av.}}$ is equivalent, up to a normalization factor,
 	to the generalized local spin imbalance
 	\cite{luitz2016extended,smith2016many,jonay2024slow} defined for arbitrary
 	half-filled $z$-basis product states: 
 	\begin{equation}
 		\mathcal I_{\mathrm{gen}}(T_0)
 		\equiv
 		\frac{4}{L}
 		\sum_j
 		\langle\psi_0|
 		\hat S_j^z(0)\hat S_j^z(T_0)
 		|\psi_0\rangle
 		=
 		2m^{\mathrm{av.}}.
 	\end{equation}}
  Similarly, we further introduce the averaged nearest-neighbor correlation ${G}^{\text{av}.}$, defined as 
   \begin{eqnarray}
   		G^{\mathrm{av.}}
   		&=&
   		\frac{1}{L-1}
   		\sum_{j}
   		e^{i\beta_j}
   		\biggl(
   		\langle \psi_{\mathrm{ini}}|
   		\hat{S}^z_j \hat{S}^z_{j+1}
   		|\psi_{\mathrm{ini}}\rangle
   		\nonumber\\
   		&&\qquad
   		-
   		\langle \psi_{\mathrm{ini}}|
   		\hat{S}^z_j
   		|\psi_{\mathrm{ini}}\rangle
   		\langle \psi_{\mathrm{ini}}|
   		\hat{S}^z_{j+1}
   		|\psi_{\mathrm{ini}}\rangle
   		\biggr).
   	\end{eqnarray}
  with $\beta_j\in \{0,\pi\}$ which guarantees that{ $$e^{i\beta_j}\big(\langle\psi_{0^+}| \hat{S}^z_j\hat{S}^z_{j+1}|\psi_{0^+}\rangle-\langle\psi_{0^+}|\hat{S}^z_{j}|\psi_{0^+}\rangle\langle\psi_{0^+}|\hat{S}^z_{j+1}|\psi_{0^+}\rangle\big) > 0.$$
  Here $|\psi_{0^+}\rangle\equiv\lim_{T_0\to{0^+}}e^{-i\hat{H}_{\text{Th}}T_0}|\psi_0\rangle$.}
  The numerical results shown in Fig.~\ref{fig:num3}(a) clearly demonstrate the relaxation of the local magnetization in the first dynamical regime. On the other hand, inter-site correlations remain finite in the second regime, leading to a further increase in the saturation entanglement. We further estimate the saturation value of the WRE based on LIOMs. We assume that the long-time state can be described by a generalized Gibbs ensemble:
  \begin{equation}
  \rho_\text{LIOM} = \frac{1}{Z} \exp\Big(\sum_j\lambda_j \hat{S}_j^z + \sum_{j<k}\xi_{jk} \hat{S}_j^z \hat{S}_{k}^z +\cdots\Big),
  \end{equation}
  Here, $\cdots$ denotes higher-body terms, and $Z$ is the corresponding partition function that guarantees $\mathrm{tr}[\rho_{\text{LIOM}}]=1$. For simplicity, we retain only nearest-neighbor couplings $\xi_{j,j+1}$ and neglect higher-order terms. We first numerically evaluate the local magnetizations and nearest-neighbor spin correlations in the structured initial state $|\psi_{\mathrm{ini}}\rangle$. For each initial state, the parameters $\lambda_j$ and $\xi_{j,j+1}$ in $\rho_{\mathrm{LIOM}}$ are determined by solving the matching equations
\begin{equation}
\begin{aligned}
\langle \psi_{\mathrm{ini}} |
\hat{S}^z_j
| \psi_{\mathrm{ini}} \rangle
&=
\operatorname{Tr}\!\left(
\rho_{\mathrm{LIOM}} \hat{S}^z_j
\right),
\\
\langle \psi_{\mathrm{ini}} |
\hat{S}^z_j \hat{S}^z_{j+1}
| \psi_{\mathrm{ini}} \rangle
&=
\operatorname{Tr}\!\left(
\rho_{\mathrm{LIOM}} \hat{S}^z_j \hat{S}^z_{j+1}
\right).
\end{aligned}
\end{equation}
We then compute the saturation value of the second R\'enyi entropy for a given subsystem $A$ using the density matrix $\rho_{A,\text{LIOM}}=\text{tr}_B [\rho_\text{LIOM}]$.  For this generalized Gibbs ensemble, our entropy definition uses the larger of the two subsystem purities, 
  \begin{equation}
  e^{-S^{(2)}_{A,\text{LIOM}}}=\text{max}\big\{\text{tr}[\rho_{A,\text{LIOM}}^2], \text{tr}[\rho_{B,\text{LIOM}}^2]\big\}.
  \end{equation}
  This leads to the green line in Fig.~\ref{fig:num3}(b). The remaining deviations contain contributions from long-range and higher-body correlations neglected in the effective description. 

  Nevertheless, the Gibbs ensemble contains thermal entropy, which leads to deviations at large initial-state entanglement. To remedy this issue, we replace the generalized Gibbs ensemble $\rho_{\text{LIOM}}$ with its pure-state realization, namely the Scrooge (random-state) ensemble \cite{jozsa1994lower}. The Scrooge ensemble is obtained by drawing pure states, $ |\psi_\phi\rangle \equiv \sqrt{\rho_{\text{LIOM}}}\,|\phi\rangle/\sqrt{\langle \phi|\rho_{\text{LIOM}}|\phi\rangle}$
with probability density
$p_\rho(\phi) = D\langle \phi|\rho_{\text{LIOM}}|\phi\rangle$, where $|\phi\rangle$ is sampled from the Haar ensemble and $D$ is the Hilbert space dimension. In this case, the WRE of the Scrooge ensemble is given by 
\begin{equation}
	\overline{S_W^{(2)}}=
	\frac{\sum_\phi p_\rho(\phi)\,
		S_W^{(2)}(|\psi_\phi\rangle)}
	{\sum_\phi p_\rho(\phi)},
\end{equation}
as shown by the black line in Fig.~\ref{fig:num3}(b). The results show an improved prediction for the growth of WRE, suggesting that the Scrooge ensemble provides a good description of pure-state ensembles that retain the local properties of the generalized Gibbs ensemble.

  \begin{figure}[t]
    \centering
    \includegraphics[width=1\linewidth]{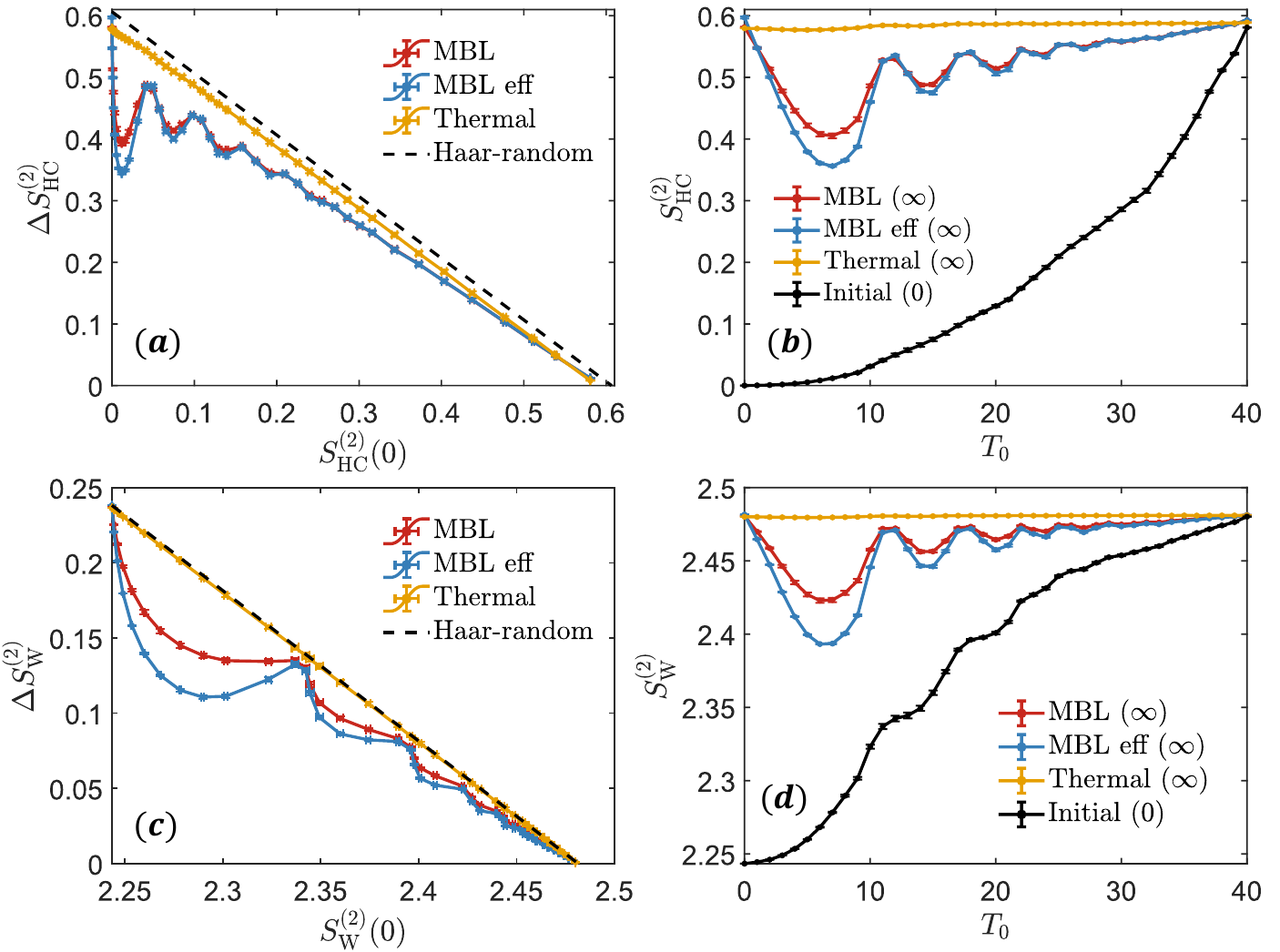}
    \caption{ The total entanglement growth of both the HCEE and the WRE is shown for different Hamiltonians and initial states at system size {$L = 14$}, where we choose $|\psi_0\rangle=|+x,+y,\cdots,+x,+y\rangle$. Despite the oscillations observed in the MBL regime, the saturation values for both the MBL and thermal cases closely match those of Haar random states, as shown in the black dashed lines. { All results are obtained by averaging over 50 independent realizations of the random fields.} }
    \label{fig:num4}
  \end{figure}

  Our analysis further suggests that the nonmonotonic behavior of the total entanglement growth depends crucially on the choice of $|\psi_0\rangle$. If we instead consider product states with vanishing local magnetization, the first regime would not be expected to emerge. The numerical results for $|\psi_0\rangle=|+x,+y,\cdots,+x,+y\rangle$, chosen to guarantee a vanishing energy expectation value ${\langle \psi_0 | \hat{H}_t | \psi_0 \rangle} = 0$, are presented in Fig.~\ref{fig:num4}. {This state satisfies $\langle \hat S_i^z\rangle=0$ at every site and hence
  	$\langle \hat S_{\rm tot}^z\rangle=0$, where
  	$\hat S_{\rm tot}^z=\sum_i \hat S_i^z$.
  	However, it is not an eigenstate of $\hat S_{\rm tot}^z$; instead, it is a coherent superposition of different total-$S^z$ sectors, whose weights are separately conserved during the evolution.} While oscillations emerge in MBL systems for both the microscopic and phenomenological models, the overall growth of total entanglement stands in sharp contrast to Fig.~\ref{fig:num2}. In particular, no separation into distinct dynamical regimes occurs, in agreement with our theoretical predictions. Collectively, our results provide a detailed dissection of how local conserved quantities encode initial-state information and govern long-time entanglement saturation.

  \section{Discussions}
  In this work, we investigate the initial-state dependence of entanglement dynamics in MBL systems, taking both the random-field XXZ model and a phenomenological model as examples, alongside a comparison with the thermalized case. We employ structured initial states prepared by evolving a product state under a chaotic thermal Hamiltonian and characterize entanglement using the second R\'enyi entropy (HCEE, $n$QEE) and the Wehrl-R\'enyi entropy (WRE). We first demonstrate that the evolution of the WRE closely tracks that of the $n$QEE, highlighting the dominance of the entanglement buildup process. We further examine the non-monotonic growth of total entanglement in MBL systems for $z$-direction product states, which can be divided into two distinct regimes based on the entanglement of the initial state. The first regime is governed by the relaxation of LIOMs, which amplifies entanglement growth by expanding the accessible Hilbert-space dimension. The second regime is dominated by inter-site correlations, wherein the MBL phenomenological model aligns with the microscopic XXZ model. These findings are validated through analyses of {local} magnetization relaxation, persistent inter-site correlations, and the absence of the first regime for product states polarized along $x-$ or $y-$direction. 

  We conclude with several remarks on future research directions. It would be worthwhile to extend this work to explore initial-state dependence in systems exhibiting weak ergodicity breaking, such as those with quantum many-body scars \cite{moudgalya2018entanglement,turner2018weak,choi2019emergent,ho2019periodic,lin2019exact,turner2021correspondence,serbyn2021quantum,yao2022quantum,su2023observation}. Generalizing the analysis to structured mixed states with tunable quantum entanglement and thermal entropy may further clarify how coupling to an environment influences multipartite entanglement. Finally, establishing connections between the observed initial-state dependence and quantum information applications, such as quantum metrology \cite{giovannetti2006quantum,giovannetti2011advances}, could provide practical insights for the advancement of quantum science and technology.

  \vspace{5pt}
  \textit{Acknowledgement.} This project is supported by the Shanghai Rising-Star Program under grant number 24QA2700300 (P. Zhang), the NSFC under grant 12374477 (P. Zhang), the Quantum Science and Technology-National Science and Technology Major Project 2024ZD0300101 (P. Zhang),  the Xuemin Institute of Advanced Studies, Fudan University (P. Zhang), the Shanghai Qiguang Natural Science Development Foundation (P. Zhang), and the Outstanding Innovative Talents 
  Cultivation Funded Programs 2026 of Renmin University of China (C. Xu).

\appendix

\appendix
{
\section{Robustness to Parameters of the Phenomenological Model}
\label{parameter}

In the main text, we employ a simplified phenomenological model
to provide a qualitative comparison with the microscopic MBL dynamics.
Starting from the phenomenological model (\ref{Heff}),
we approximate the LIOMs by the bare spin operators,
$\hat{Q}_i=\hat{S}_i^z$, and retain only the two-body interaction terms.
The resulting Hamiltonian reads
\begin{equation}
	\hat{H}_{\rm eff}
	=
	\sum_{i=1}^{L} g_i \hat{S}_i^z
	+
	\sum_{i<j}
	\tilde{J}_{ij}e^{-|i-j|/\xi}
	\hat{S}_i^z\hat{S}_j^z ,
	\label{eq:Heff_app}
\end{equation}
where $\tilde{J}_{ij}$ and $g_i$ are independently sampled from
uniform distributions in $[-J_0,J_0]$ and $[-g_0,g_0]$, respectively.

\begin{figure}[t]
	\centering
	\includegraphics[width=1\linewidth]{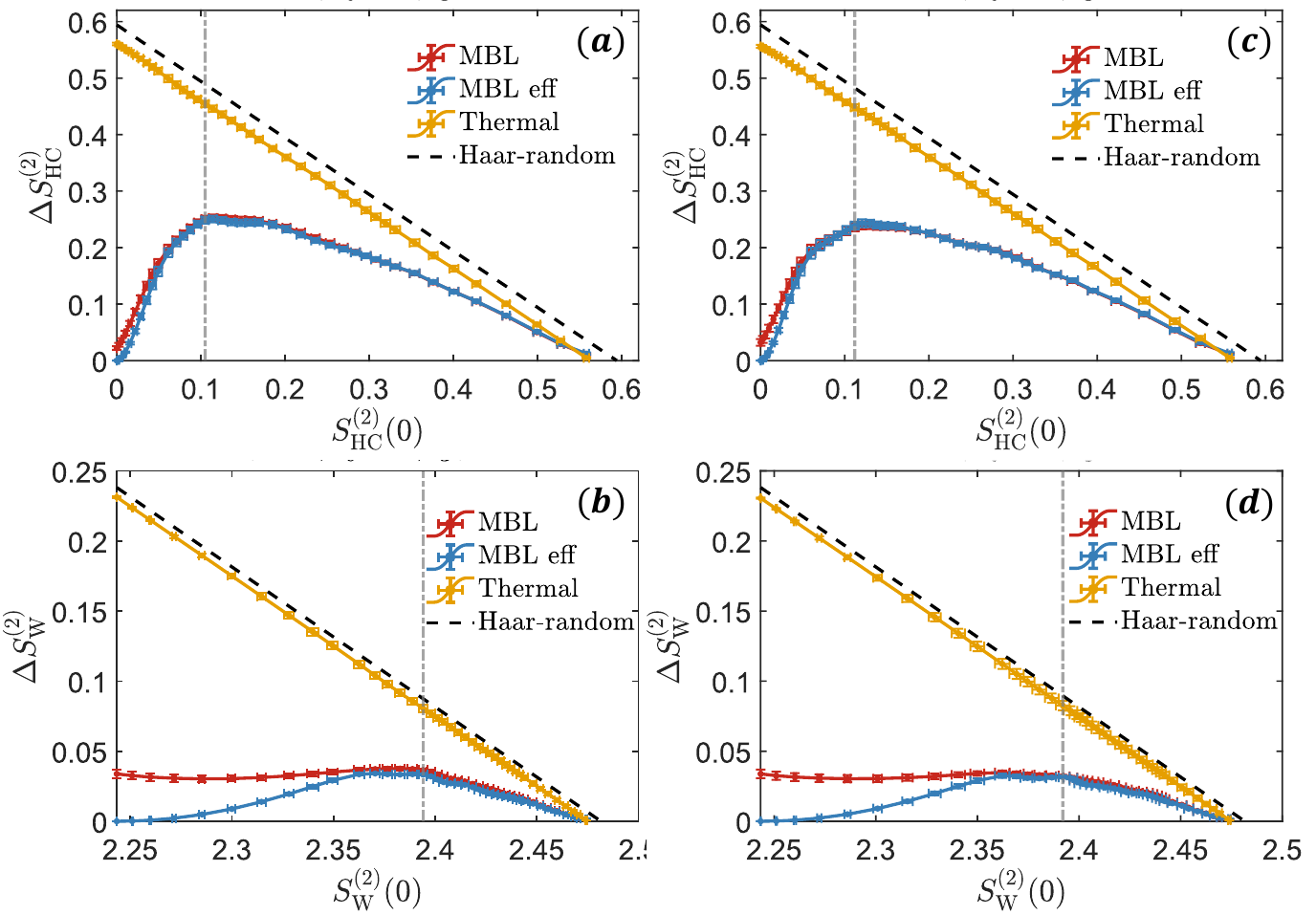}
\caption{
	Net entanglement growth of the HCEE and WRE for two additional parameter choices of the phenomenological model at system size $L=14$: (a,b) $(g_0,\xi)=(1,1)$ and (c,d) $(g_0,\xi)=(0,2)$, with $J_0=1$. The reference parameter set $(g_0,\xi)=(0,1)$ is used in the main text. Except for the parameters of the phenomenological model, all other parameters and the regime boundaries are the same as those in Fig.~\ref{fig:num2}. All results are averaged over 50 independent realizations.}
	\label{fig:parameter}
\end{figure}

The replacement $\hat{Q}_i=\hat{S}_i^z$ should be understood as a
strong-disorder approximation rather than an exact representation of
the microscopic MBL Hamiltonian. In the limit $W\rightarrow\infty$,
the LIOMs approach the bare local spin operators, whereas at large but
finite disorder they acquire quasi-local dressing. Likewise, the
omission of higher-body LIOM interactions constitutes a truncation of
the generic LIOM expansion. 
Within the truncated phenomenological model, all terms in Eq.~(\ref{eq:Heff_app}) commute,
so that the time-evolution operator can be written as
\begin{equation}
	e^{-i\hat{H}_{\rm eff}T}
	=
	\left[
	\prod_i e^{-ig_i\hat{S}_i^zT}
	\right]
	\exp\left[
	-iT\sum_{i<j}
	J_{ij}\hat{S}_i^z\hat{S}_j^z
	\right].
	\label{eq:Ueff_app}
\end{equation}
The first factor consists only of on-site unitary rotations and therefore
does not generate entanglement. Consequently, changing $g_0$ does not
affect the HCEE or the WRE considered here. By contrast, the localization
length $\xi$ controls the spatial range of the effective LIOM interactions
and can modify the associated dephasing timescale.

The parameter $J_0$ sets the overall interaction-energy scale and is therefore fixed to $J_0=1$. In the main text, the reference parameters are chosen as
\begin{equation}
	(g_0,\xi)=(0,1).
\end{equation}
To examine the robustness of our conclusions against these parameter
choices, here we additionally consider
\begin{equation}
	(g_0,\xi)=(1,1)
	\qquad\text{and}\qquad
	(g_0,\xi)=(0,2).
	\label{eq:parameter_sets}
\end{equation}
As shown in Fig.~\ref{fig:parameter}, both additional parameter choices
exhibit the same qualitative dependence on the preparation time $T_0$
as the reference case presented in the main text. In particular,
increasing the interaction range from $\xi=1$ to $\xi=2$ does not remove
the two-regime behavior. The result for $g_0=1$ is also consistent with
the fact that the single-LIOM terms do not affect the entanglement measures considered here.
These results indicate that the qualitative conclusions drawn from the
phenomenological model are robust within the parameter range tested.
We emphasize that this simplified model is used only as a qualitative
reference for the microscopic MBL dynamics and is not intended as a
quantitatively complete representation of the dressed LIOMs or of the
higher-body terms in the full effective MBL Hamiltonian.

}

\bibliography{ref.bib}

\end{document}